# Two-Dimensional Inversion Asymmetric Topological Insulators in Functionalized III-Bi Bilayers

Yandong Ma[†][*], Liangzhi Kou[‡], Binghai Yan[§][//], Chengwang Niu[⊥], Ying Dai[⊥], and Thomas Heine[†][*]

[†] Engineering and Science, Jacobs University Bremen, Campus Ring 1, 28759 Bremen, Germany
[‡] School of Chemical Engineering, University of New South Wales, Sydney, NSW 2052, Australia
[§] Max Planck Institute for Chemical Physics of Solids, Noethnitzer Str. 40, 01187 Dresden, Germany
[//] Max Planck Institute for the Physics of Complex Systems, Noethnitzer Str. 38, 01187 Dresden, Germany
[⊥] School of Physics, Shandong University, Shandanan Str. 27, 250100 Jinan, People's Republic of China

*Corresponding author: myd1987@gmail.com (Y.M.); t.heine@jacobs-university.de (T.H.)

The search for inversion asymmetric topological insulators (IATIs) persists as an effect for realizing new topological phenomena. However, so for only a few IATIs have been discovered and there is no IATI exhibiting a large band gap exceeding 0.6 eV. Using first-principles calculations, we predict a series of new IATIs in saturated Group III-Bi bilayers. We show that all these IATIs preserve extraordinary large bulk band gaps which are well above room-temperature, allowing for viable applications in room-temperature spintronic devices. More importantly, most of these systems display large bulk band gaps that far exceed 0.6 eV and, part of them even are up to ~1 eV, which are larger than any IATIs ever reported. The nontrivial topological situation in these systems is confirmed by the identified band inversion of the band structures and an explicit demonstration of the topological edge states. Interestingly, the nontrivial band order characteristics are intrinsic to most of these materials and are not subject to spin-orbit coupling. Owning to their asymmetric structures, remarkable Rashba spin splitting is produced in both the valence and conduction bands of these systems. These predictions strongly revive these new systems as excellent candidates for IATI-based novel applications.

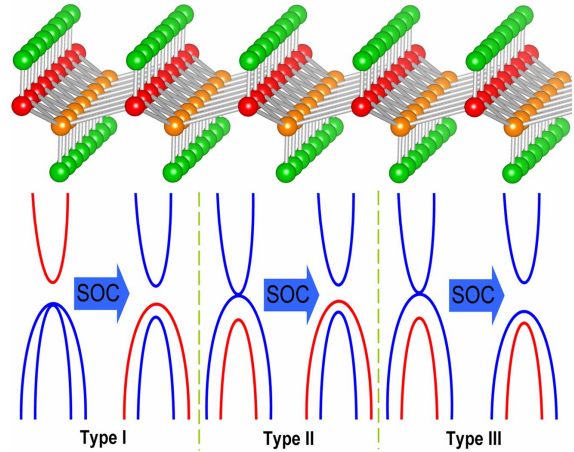

TOC Figure

## I. Introduction

Topological insulators (TIs) are a new class of materials which reveals a fundamental interplay between the electronic structure and topology, it provides a platform for potentially promising applications in spintronics.[1,2] This novel quantum matter can support odd number of gapless boundary states inside a bulk band gap, which are topologically protected and robust against perturbation. So far, a number of compounds have been theoretically or experimentally confirmed to be TIs.[3-20] In practical applications, TIs with a large bulk band gap are more desirable because of their ability to stabilize the edge current against the interference of thermally activated carriers in the bulk. In order to obtain practical TIs with preferable large bulk band gaps, extensive theoretical and experimental efforts have been devoted to Bi-related materials in view of the large atomic spin-orbital coupling (SOC) strength in Bi atom. Examples include $Bi_2Se_3$,[7] BiTeCl,[8] Bi(111) bilayer,[9] bilayers of group III elements with Bi,[10] $BaBiO_3$,[11] chemically modified Bi honeycomb lattices,[12,13] as well as $Bi_4Br_4$ thin films[14].

Depending on their crystal structures, TIs can be classified into inversion symmetric TIs (ISTIs) and inversion asymmetric TIs (IATIs). Although the presence of inversion symmetry is helpful in identifying TIs owing to the existence of the parity criterion,[21,22] the IATIs arise intense attention by their perfect performance in realizing new topological phenomena in practical materials. It is due to the inversion symmetry break that IATIs can host many nontrivial phenomena such as crystalline-surface-dependent topological electronic states,[9,23] pyroelectricity,[24] and natural topological p–n junctions,[25] as well as the realization of topological magneto-electric effects.[26,27] In addition, the occurrence of superconductivity in IATIs would give raise to unique features such as a large upper critical field beyond the Pauli limit[28,29] and topological superconductivity with the Majorana edge channels[30]. All these properties lead to a great potential of IATIs in device paradigms for spintronics and quantum information processing. Despite the importance of IATIs, up to now, few materials have been shown to be IATIs[8,10,22,31,32]. In addition, no IATI with a bulk band gap exceeding 0.6 eV has ever been discovered either theoretically or experimentally.[8,10,22,31,32] The small bulk band gaps of the obtained IATIs severely limit the working environments, in particular real-world applications are restricted due to the need to operate at cryogenic temperature. Therefore, the search for new IATIs with large bulk band gaps is urgently required.

Here, on the basis of first-principles calculations, we report on a series of two-dimensional (2D) IATIs with significantly large bulk band gaps in hydrogenated and fluorinated Group III-Bi, which we call III-Bi in the remainder for the sake of simplicity, (H2/F2-GaBi, H2/F2-InBi and H2/F2-TlBi) bilayers. The topological characteristics of these systems are confirmed by the identification of band inversion and the explicit presence of the nontrivial topological edge states. More remarkably, except hydrogenated GaBi and InBi, all the bulk band gap of these IATIs exceed 0.6 eV and, part of them are even approach to about 1 eV, making them viable for high temperature applications. Furthermore, our results suggest that, except for hydrogenated GaBi and InBi, the nontrivial band order characteristics of other materials are intrinsic regardless of SOC. Together with the excellent nontrivial topological characteristics in these systems, we establish that these IATIs can produce remarkable Rashba spin splitting, deriving from their inversion asymmetry-induced strong polar field. These findings make these new IATIs promising platforms for unusual topological phenomena and possible applications at high temperature.

**II. Computational Details**

Our first-principles calculations are performed using the Vienna *ab initio* simulation package (VASP).[33,34] The projector-augmented wave (PAW) method[35] is used to describe the electron-ion potential. The exchange-correlation potential is approximated by the generalized gradient approximation (GGA) in the Perdew-Burke-Ernzerhof (PBE) form.[36] A plane-wave basis set with kinetic energy cutoff of 500 eV is used. Periodic boundary conditions are employed to simulate these 2D systems. Vacuum space between a constructed structure and its periodic mirrors is chosen to be no less than 18 Å, which is sufficient for energy convergence. We employ a k-point set generated by the 17×17×1 Monkhorst-Pack mesh[37] for both geometry optimizations and self-consistent calculations. The atomic coordinates of all atoms in the unit cell and the cell's length are fully relaxed, with the forces on every atom converged to within 0.01 eV/ Å. In the self-consistent calculations, SOC is included. For the phonon dispersion relations, the finite displacement method as implemented in the CASTEP code[38,39] and PBE for the exchange-correlation are used.

## III. Results and Discussion

H2/F2-III-Bi possesses a graphane-type crystal structure and a hexagonal symmetry with four atoms in one unit cell. Within each unit, the H/F atom, III atom, Bi atom and H/F atom are stacked along the hexagonal axis, as shown in **Fig. 1a**. The stacking order of the layers of the three atomic species breaks the inversion symmetry. Due to the absence of inversion symmetry, the charge in H2/F2-III-Bi is expected to distribute unevenly along the hexagonal axis, giving rise to intrinsic surface dipole moment (ranging from 0.065 to 0.827 Debye per unit cell). Together with the 2D configuration, the SOC can result in a spin splitting of the energy bands of H2/F2-III-Bi away from the time-reversal invariant momenta (TRIM), which has the nature of the so-called Rashba effect,[40] as we will show. The optimized lattice constants of H2-GaBi, H2-InBi, H2-TlBi, F2-GaBi, F2-InBi, and F2-TlBi are 4.588, 4.891, 4.985, 4.846, 5.160, and 5.286 Å, respectively, with the buckling height of III-Bi layer being 0.795, 0.851, 0.872, 0.502, 0.449, and 0.297 Å. The dynamic stability of the H2/F2-III-Bi thin films is confirmed by the phonon dispersion curves which are shown in **Fig.1** and **Fig. S1**. All branches of the phonon dispersion curves have positive frequencies and no imaginary phonon modes are found, confirming the stability of these thin films.

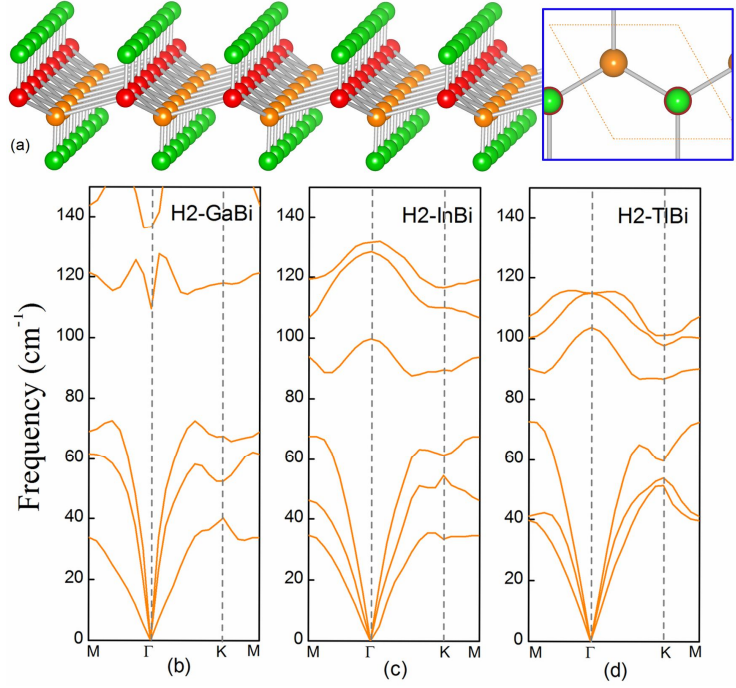

**Figure 1**. (a) Geometric structure of H2/F2-III-Bi thin films. The H/F, III, and Bi atoms are highlighted in green, orange, and red spheres, respectively. Phonon band dispersion relations calculated for (b) H2-GaBi, (c) H2-InBi, and (d) H2-TlBi thin films.

In **Fig. 2**, we show the electronic band structures of H2/F2-III-Bi thin films in the absence SOC. Without considering the SOC effect, the band structures of H2-GaBi and H2-InBi show a semiconductor nature with a direct band gap located at Γ point (see **Fig. 2a** and **2c**). By projecting the bands onto different atomic orbitals we find that the energy spectrum near the Fermi level at the Γ point mainly comes from one s and two p orbitals (excluding spin), which show "–" and "+" parities, respectively. At the Γ point, the two p orbitals are energy-degenerate (excluding spin), while the bands away from the Γ point are well separated. In H2-GaBi and H2-InBi, the Fermi level is located between the s orbital and the two p orbitals, rendering the "–" parity above the two "+" parity states, as illustrated in **Fig. 2a** and **2c**. Consequently, H2-GaBi and H2-InBi shows a normal band order in the absence of SOC. Compared with H2-GaBi and H2-InBi, the band structures of F2-GaBi and F2-InBi are drastically modified, as shown in **Fig. 2g** and **2i**. We can find that F2-GaBi and F2-InBi are gapless semiconductors when SOC is not considered, also referred as semimetal, with the valence band maximum and conduction band minimum being degenerate at the Fermi level. In contrast to the normal order in H2-GaBi and H2-InBi, the band order at the Γ point is inverted in F2-GaBi and F2-InBi. Specifically, the "−" parity state is shifted below the two "+" parity states, while the band orders at the other TRIM are unchanged. As observed in previous works,[41] inversion of bands with opposite parities is a strong indication of the formation of topologically nontrivial phases. This suggests that F2-GaBi and F2-InBi could be IATIs once upon opening an energy gap at the touching point. The modification of the band order upon different surface functionalization is directly related to the relation of bond length and orbital splitting. In the case without SOC, owning to the chemical bonding among the atoms, the states near the Fermi level are split into bonding and antibonding states, which we denote as $|s^{\pm}\rangle$ and $|p_{x,y}^{\pm}\rangle$, with the subscript ± representing the parity. For H2/F2-GaBi/InBi, the bands near the Fermi level are mainly derived from the $|s^{-}\rangle$ and $|p_{x,y}^{+}\rangle$ orbitals, with the $|p_{x,y}^{+}\rangle$ orbital lies below the $|s^{-}\rangle$ orbital. By comparing the lattice constants of the systems mentioned above one may notice that the lattice constant of F2-GaBi/InBi is larger than that of H2-GaBi/InBi. The larger lattice constant in F2-GaBi/InBi leads to a weaker s-p hybridization, and accordingly a smaller energy separation between the bonding and antibonding states. Consequently, compared with that of H2-GaBi/InBi, the $|s^{-}\rangle$ orbital is downshifted while the $|p_{x,y}^{+}\rangle$ orbital is upshifted in F2-GaBi/InBi, yielding the fact that the $|s^{-}\rangle$ orbital lies below the $|p_{x,y}^{+}\rangle$ orbital. Noteworthy, for normal insulators, the $|s^{-}\rangle$ orbital typically lies above the $|p_{x,y}^{+}\rangle$ orbital, the band order is thus obviously inverted in F2-GaBi/InBi. In addition, because of the lattice symmetry and the absence of SOC, the $|p_{x,y}^{+}\rangle$ orbital are energy degenerate (excluding spin), which thus lead to that fact that F2-GaBi and F2-InBi are gapless semiconductors, see **Fig. 2g** and **2i**. The similar situation in F2-GaBi and F2-InBi is observed in H2-TlBi and F2-TlBi. To further confirm these results, we have employed the more sophisticated Heyd−Scuseria−Ernzerhof hybrid functional method (HSE06)[42] to calculate the band structures of these systems and the corresponding results are plotted in **Fig. S2**. The results showed **Fig. S2** are in good agreement with these obtained here.

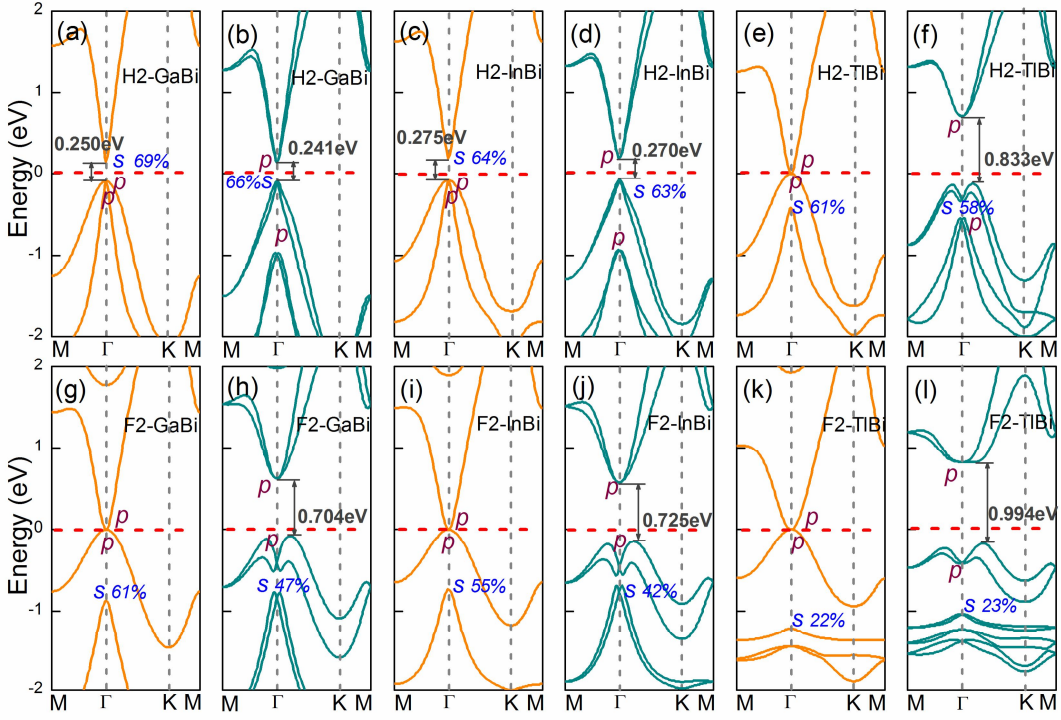

**Figure 2**. The electronic band structures of (a and b) H2-GaBi, (c and d) H2-InBi, (e and f) H2-TlBi, (g and h) F2-GaBi, (i and j) F2-InBi, and (k and l) F2-TlBi thin films. The orange line and dark cyan line represent the band structures without and with SOC, respectively. The horizontal dashed lines indicate the Fermi level. Insert: the value of the band gap induced by SOC; the main projection of the bands near the Fermi level at Γ point and the corresponding projection weight.

The strong SOC from Bi is expected to have significant impact on the electronic band structures of H2/F2-III-Bi thin films, which is confirmed in **Fig. 2**. In H2-GaBi and H2-InBi, the SOC effect lifts the energy-degeneracy of the valence band maximum at the Γ point, see **Fig. 2b** and **2d**. As a result, one of the degenerated valence bands is upshifted, while the conduction band is downshifted. By considering the orbital contribution to the band, one can find that, with switching on SOC, two bands with different parities are inverted around the Fermi level in H2-GaBi and H2-InBi. The band inversion strongly points to the inversion asymmetric nontrivial topological phases in H2-GaBi and H2-InBi. More interestingly, these two systems exhibit sizeable nontrivial bulk band gaps, with the global band gap of 0.241 and 0.270 eV, respectively, indicating that the QSH effect in these two systems can be easily observed at room-temperature. Along with the band inversion, the band structures of H2-GaBi and H2-InBi experience another significant modification, namely the spin-splitting of bands. It is well known that the structure inversion symmetry will keep the spin-degeneracy of energy bands as long as the time-reversal symmetry is kept. Otherwise, the spin degeneracy of the bands would be lifted at the generic k points by SOC in systems with inversion asymmetry, which can be well exemplified by the Rashba effect.[40] As described above, the particular geometric structures of these systems break the inversion symmetry of the crystal potential, which would give raise to the Rashba spin splitting of the bulk bands in these systems. As shown in **Fig. 2b** and **2d**, the valence band maximum of these systems locates slightly off the Γ point, thus forming a significant Rashba spin splitting in these systems. Besides, to show the strength of SOC in these systems more intuitively, we label the SOC-induced band gap between the two p orbitals around the Fermi level at the Γ point as $E_{SOC}$, to be distinguished from the conventional insulating band gaps. The SOC strength $E_{SOC}$ is about 1.125 and 1.132 eV, respectively, for H2-GaBi and H2-InBi.

As it can be seen in **Fig. 2h** and **2j**, the inclusion of SOC also disrupts the degeneracy of the two bands around the Fermi level in F2-GaBi and F2-InBi, with the valence band and conduction band shifting downwards and upwards, respectively. For the sake of the same band components (i.e., mainly contributed by the p orbital) of these two bands at the Γ point, the SOC-driven band inversion around the Fermi level cannot occur in F2-GaBi and F2-InBi. Hence, the nontrivial band orders in F2-GaBi and F2-InBi are unchanged upon considering SOC. F2-GaBi and F2-InBi are therefore also nontrivial 2D IATIs, similar to H2-GaBi and H2-InBi. Of course more proofs are needed, as we will show below by the demonstration of the topological edge states of these systems. An interesting feature of F2-GaBi and F2-InBi is that the effect of SOC is only producing an energy gap at the touching point around the Fermi level, but not inducing band inversion as in H2-Ga(In)Bi (namely SOC is not relevant for the formation of nontrivial band order). In fact, a similar situation is observed in well-known 2D TIs like graphene[43] and silience[44], where the inclusion of SOC also does not change the band order between the valence and conduction bands at the TRIM points.

Therefore, similar to graphene[43] and silience[44], the bands of F2-GaBi and F2-InBi at the TRIM points of these systems display "intrinsic nontrivial band order" regardless of SOC, different from the TIs where the driving force of band inversion is SOC. Similar nontrivial topological characteristics are found in H2-TlBi and F2-TlBi, see **Fig. 2f** and **2l**. By looking at the band dispersions near the Fermi level of these systems in the presence of SOI, the bands also display Rashba spin splitting, see **Fig. 2**.

Besides these particular properties, another remarkable common feature of F2-GaBi, F2-InBi, F2-TlBi and H2-TlBi is their extraordinary large bulk band gap. In detail, the global nontrivial band gaps of F2-GaBi, F2-InBi, F2-TlBi and H2-TlBi are 0.704, 0.725, 0.994 and 0.833 eV, respectively, which are sufficiently large for practical application at high temperature. It is important to note that no IATI with a bulk band gap exceeding 0.6 eV has ever been discovered either theoretically or experimentally up to now. Therefore, such giant band gaps in F2-GaBi, F2-InBi, F2-TlBi and H2-TlBi considerably exceed the upper limit of the band gap in the field of IATIs. These band gap magnitudes are advantageous for stabilizing the edge current against the interference of thermally activated carriers in the bulk. In addition to the small size of the bulk band gap, another major obstacle in the field of IATIs is the material shortage: only few materials have been shown to be IATIs.[8,10,23,31,32] In contrast, we predict all these III-Bi materials are IATIs. The finding of such a series of IATIs with large band gaps is not only of technological interest but also of fundamental significance in spintronics. The large nontrivial band gaps of these materials originate form the strong SOC of the σ orbitals around the Fermi level, with the $E_{SOC}$ of 1.029, 1.042, 1.235 and 1.278 eV, respectively, for F2-GaBi, F2-InBi, F2-TlBi and H2-TlBi.

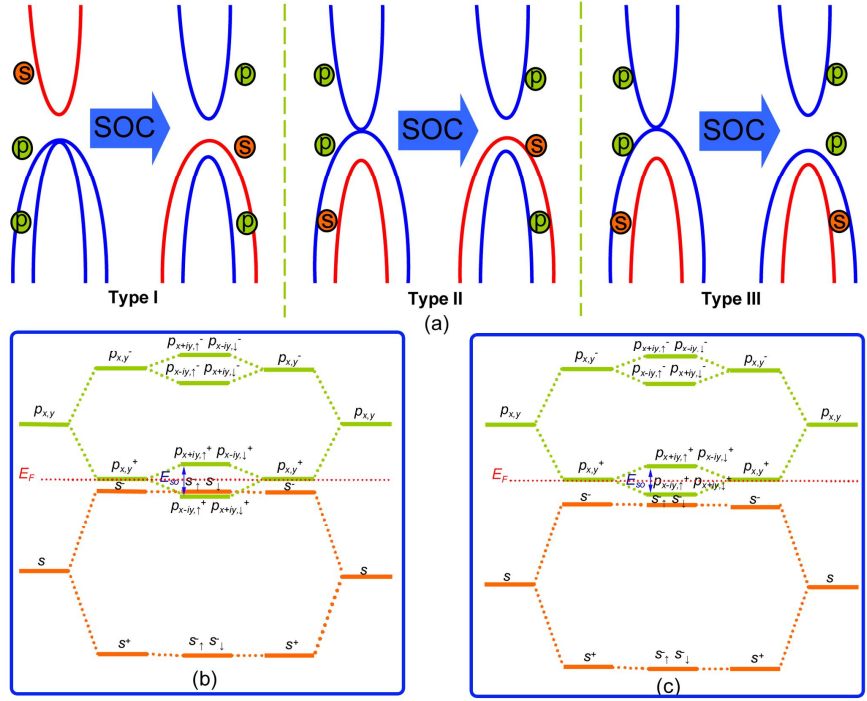

**Figure 3**. (a) Illustration of the effect of SOC on the bands around the Fermi level in H2/F2-III-Bi thin films. The evolution of atomic s and $p_{x,y}$ orbitals of (b) H2-TlBi and (c) F2-TlBi into the band edges at Γ point is described as the crystal field splitting and SOC are switched on in sequence. The horizontal dashed lines in (b) and (c) indicate the Fermi level.

When checking the trend of nontrivial gap variation in the studied cases, we can find one obvious feature. For all these systems, the nontrivial band gap of F2-III-Bi is larger than that of its hydrogenated counterpart. Especially for H2/F2-GaBi (and H2/F2-InBi), the nontrivial band gap of F2-GaBi (and F2-InBi) is even three times (and two times) larger than that of H2-GaBi (and H2-InBi). However, one can find that, in sharp contrast to the significant difference in the nontrivial band gaps, the SOC strength $E_{SOC}$ in these systems remains almost unchanged when one changes hydrogenation to fluorination. Such a discrepancy stems from their particular bands around the Fermi level. To elucidate this discrepancy, we show in **Fig. 3a** the illustration of the effect of SOC on the bands around the Fermi level of H2/F2-III-Bi. It can be seen there are three scenarios for the effect of SOC on the bands around the Fermi level: H2-GaBi and H2-InBi belong to type I, H2-TlBi belongs to type II, while F-III-Bi belongs to type III. As **Fig. 3a** illustrates for type I and type II, after including SOC, the s orbital lies between rather than below the two SOC-split p orbitals, forming the valence band maximum. These results establish that, in principle, the value of the nontrivial band gap is smaller than the SOC strength $E_{SOC}$ in type I and type II. While for type III, the s orbital is well below the two p orbitals, regardless of SOC, implying the value of the nontrivial band gap is comparable to the SOC strength $E_{SOC}$. With these results in mind, one can easily understand the discrepancy mentioned above. We then take H2-TlBi and F2-TlBi as examples to get a deep insight into this feature. As schematically shown in **Fig. 3b** and **3c**, the nontrivial band gap is affected not only by SOC but also by the crystal field splitting (CFS). And the later factor is directly determined by the bond strength. Since the SOC strength is comparable for hydrogenated and fluorinated systems, the CFS is responsible for the relative position of the s orbital, as we have detailed above. This argument is corroborated by the fact that the lattice constant of F2-TlBi is larger than that of the H2-TlBi.

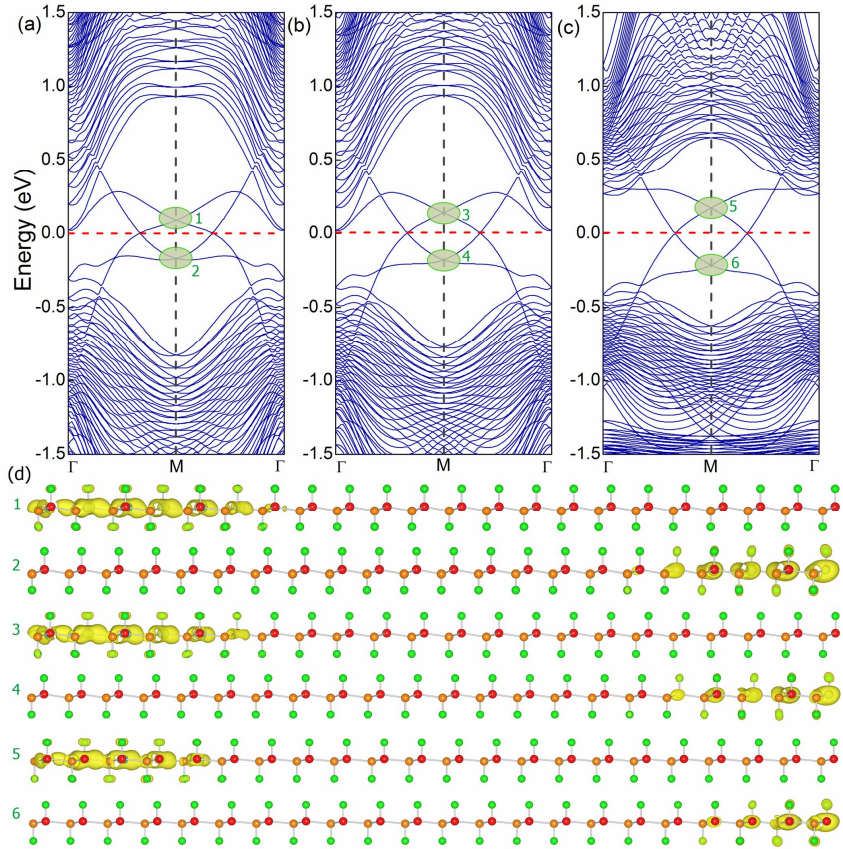

**Figure 4**. Electronic band structures for the zigzag nanoribbons of (a) F2-GaBi, (b) F2-InBi, and (c) F2-TlBi thin films. The Fermi energy is set to 0 eV. (d) Real space charge density distributions of the edge states at the Γ point for different bands, as marked in (a)-(c). The F, III, and Bi atoms are highlighted in green, orange, and red spheres, respectively.

The topologically protected conducting edge state inside the bulk band gap is the hallmark of 2D TIs. Therefore, in order to confirm the topological nontrivial nature of these systems, we have checked the existence of protected gapless edge states. They are calculated by using the slab model and by constructing a zigzag nanoribbon structure. For the zigzag nanoribbon, one side is terminated with a Bi-H/F chain and the other side is terminated with a III-H/F chain. As a consequence of this asymmetric structure, there would be two separated sets of gapless edge states for each slab, corresponding to the two opposite edges. To eliminate the coupling between their two edges, all the widths of these nanoribbons exceed 8 nm. The calculated results presented in **Fig. 4a-4c** and **Fig. S3a-S3c** clearly show that the helical edge states form bands connecting the valence and conduction bands and crossing linearly at the M point. Each system exhibits two Dirac cones (**Fig. 4**). They are split due to the asymmetric edges as indicated above, and we label the Dirac points at the edge of Bi-H/F chain with odd number, and the ones at the edge with III-H/F chain with even number. For a single pair of edge states of each system, they are described by an odd number of crossings over the Fermi level from M to Γ point. These features strongly indicate that these systems are all indeed IATIs. The real space charge density distributions of the edge states of these systems at the Γ point are depicted in **Fig. 4d** and **Fig. S3d**. We can observe that the topologically protected Dirac cones derive mainly from the region close to the ribbon edges, and with a little spreading towards the ribbon center. One remarkable feature is that, for a given edge like the side terminated with a III-H/F chain, the width of real-space distribution of the edge states decrease gradually in the order H2/F2-GaBi > H2/F2-InBi > H2/F2-TlBi. That means that the edge states become more and more delocalized in the order H2/F2-GaBi < H2/F2-InBi < H2/F2-TlBi, as a localized state in momentum space would exhibit an extended distribution in real space. We need to emphasize that the detail of the edge states of TIs may depend on the edge, but they always exist as they are protected by the nontrivial topology.

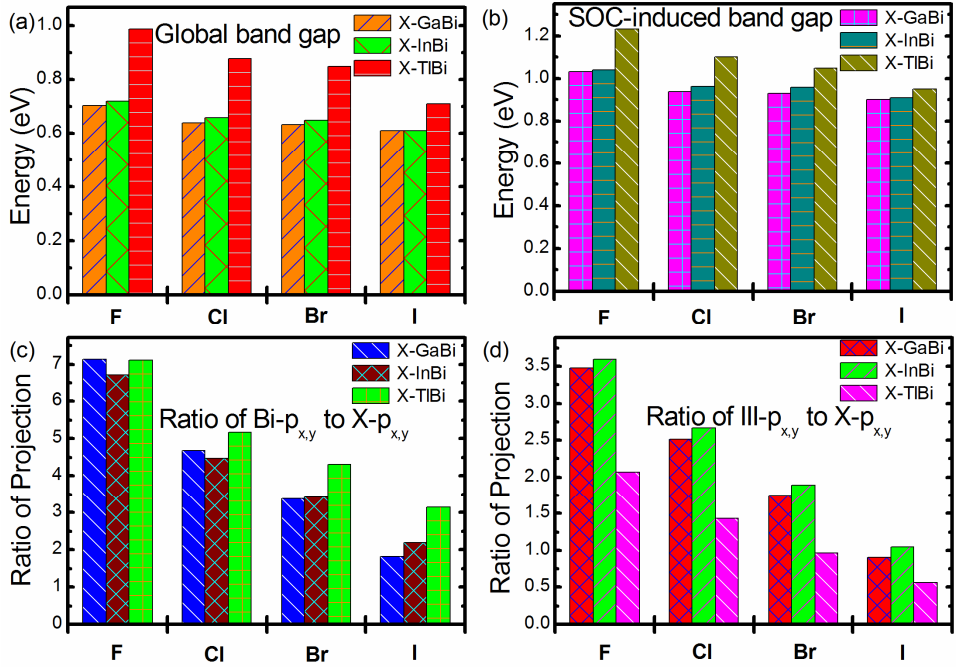

**Figure 5**. (a) The global band gaps of X2-GaBi, X2-InBi and X2-TlBi (X=F, Cl, Br and I). (b) The SOC-induced band gap around the Fermi level at the Γ point $E_{SOC}$. (c) The ratio of Bi-$p_{x,y}$ to X-$p_{x,y}$ (the one bonded to Bi) component in the σ orbital at the Fermi level at the Γ point, when excluding SOC. (d) The ratio of Ga/In/Tl-$p_{x,y}$ to X-$p_{x,y}$ (the one bonded to Ga/In/Tl) component in the σ orbital at the Fermi level at the Γ point, when excluding SOC.

At last, we wish to point out that the hydrogenation or fluorination in III-Bi systems is not the only way to achieve the large nontrivial ITATs, the same results can be obtained by passivating the surface with other halogen atom, such as Cl, Br and I. We thus performed calculation for Cl2/Br2/I2-III-Bi to investigate their topological properties. In **Fig. S4-S6**, we show the corresponding band structures of Cl2/Br2/I2-GaBi, Cl2/Br2/I2-InBi and Cl2/Br2/I2-TlBi respectively for comparison. As illustrated, our results suggest that the band structures of these systems are all similar to F2-III-Bi, and these systems also display "intrinsic nontrivial band order" regardless of SOC. Among the three scenarios depicted in **Fig. 3a** for the effect of SOC, they all belong to the type III, similar to F2-III-Bi. More remarkably, these systems exhibit large nontrivial band gaps, which all exceed 0.6 eV and, part of them exceed 0.8 eV (see **Fig. 5a**). Therefore, these systems are also promising 2D IATIs with large nontrivial band gaps. On the other hand, we can find some interesting phenomena when comparing the values of the band gaps of these systems with each other. It is known that, from F to I, the SOC strength increases in the order F < Cl < Br < I. It is therefore supposed that the values of the band gaps should increase in the order F2-III-Bi < Cl2-III-Bi < Br2-III-Bi < I2-III-Bi. And since the global band gaps of these systems are mainly determined by the SOC strength $E_{SOC}$, this order obviously should also work for the global band gaps of these systems. Yet, the fact is just opposite. As shown **Fig. 5a** and **Fig. 5b**, the $E_{SOC}$ as well as the global band gaps of these halogenated systems decrease in the contrary order, namely, F2-III-Bi > Cl2-III-Bi > Br2-III-Bi > I2-III-Bi. Explanation of this interesting contradiction is sought into the band components of the orbital around the Fermi level at the Γ point, as its splitting induced by SOC can directly determine the $E_{SOC}$ as well as global band gap. Our results shown in **Fig. 5c** clearly reveal that the ratio of Bi-$p_{x,y}$ to X-$p_{x,y}$ (the one bonded to Bi) component in the σ orbital at the Fermi level at the Γ point decrease in the order F2-III-Bi > Cl2-III-Bi > Br2-III-Bi > I2-III-Bi. The same phenomenon is observed in another system as the ratio of Ga/In/Tl-$p_{x,y}$ to X-$p_{x,y}$ (the one bonded to Ga/In/Tl) component, see **Fig. 5d**. Bearing in mind that the Bi exhibits almost the strongest SOC strength, it can get that the larger the ratio, the larger the contribution to the orbital at the Fermi level at the Γ point, and the larger the SOC strength. Based on this fact, we can understand why the $E_{SOC}$ as well as the global band gaps of these halogenated systems decrease in the contrary order F2-III-Bi > Cl2-III-Bi > Br2-III-Bi > I2-III-Bi.

**IV. Conclusion**

In conclusion, we use first-principles calculations to identify a series of new IATIs, which possess intrinsic topologically protected edge states forming QSH systems. In particular, it is emphasized that most of these systems exhibit an extraordinary large nontrivial bulk band gap that far exceeds the gap of all current discovered 2D IATIs. We further reveal that most of these systems show "intrinsic nontrivial band order" even without considering SOC, and all these systems present interesting Rashba effect due to their asymmetric geometric structures. Details of the underlying physical mechanisms of these systems are discussed. These attractive features make these systems potentially the most promising IATIs.

**Supporting Information**

Phonon band dispersion relations calculated for F2-III-Bi thin films; electronic band structures for the zigzag nanoribbons of F2-III-Bi thin films and the corresponding real space charge density distributions of the edge states; band structures of Cl2/Br2/I2-III-Bi thin films. This material is available free of charge via the Internet at http://pubs.acs.org.

**Acknowledgement**

Financial support by the European Research Council (ERC, StG 256962) and the National Science foundation of China under Grant 11174180 are gratefully acknowledged.

**Note**

The authors declare no competing financial interest.

**References**

(1) Moore, J. E. *Nature* **2013**, *464*, 194−198.

(2) Hasan, M. Z.; Kane, C. L. *Rev. Mod. Phys.* **2010**, *82*, 3045−3067.

(3) Hsieh, D.; Qian, D.; Wray, L.; Xia, Y.; Hor, Y. S.; Cava, R. J.; Hasan, M. Z. *Nature* **2008**, *452*, 970−974.

(4) Xia, Y.; Qian, D.; Hsieh, D.; Wray, L.; Pal, A.; Lin, H.; Bansil, A.; Grauer, D.; Hor, Y. S.; Cava, R. J.; Hasan, M. Z. *Nature Phys.* **2009**, *5*, 398−402.

(5) Gehring, P.; Benia, H. M.; Weng, Y.; Dinnebier, R.; Ast, C. R.; Burghard, M.; Kern, K. *Nano Lett.* **2013**, *13*, 1179−1184.

(6) Bernevig, B. A.; Hughes, T. L.; Zhang, S.-C. *Science* **2006**, *314*, 1757−1761.

(7) Kuroda, K.; Arita, M.; Miyamoto, K.; Ye, M.; Jiang, J.; Kimura, A.; Krasovskii, E. E.; Chulkov, E. V.; Iwasawa, H.; Okuda, T.; Shimada, K.; Ueda, Y.; Namatame, H.; Taniguchi, M. *Phys. Rev. Lett.* **2010**, *105*, 076802.

(8) Chen, Y. L.; Kanou, M.; Liu, Z. K.; Zhang, H. J.; Sobota, J. A.; Leuenberger, D.; Mo, S. K.; Zhou, B.; Yang, S.-L.; Kirchmann, P. S.; Lu, D. H.; Moore, R. G.; Hussain, Z.; Shen, Z. X.; Qi, X. L.; Sasagawa, T. *Nature Phys.* **2013**, *9*, 704–708.

(9) Murakami, S. *Phys. Rev. Lett.* **2006**, *97*, 236805.

(10) Chuang, F.-C.; Yao, L.-Z.; Huang, Z.-Q.; Liu, Y.-T.; Hsu, C.-H.; Das, T.; Lin, H.; Bansil, A. *Nano Lett.* **2014**, *14*, 2505−2508.

(11) Yan, B. H.; Jansen, M.; Felser, C. *Nature Phys.* **2013**, *9*, 709–711.

(12) Liu, C.-C.; Guan, S.; Song, Z.; Yang, S. A.; Yang, J.; Yao, Y. **2014**, arXiv:1402.5817.

(13) Ma, Y. D.; Dai, Y.; Kou, L. Z.; Frauenheim, T.; Heine. T. *Nano Lett.* **2015**, DOI: 10.1021/nl504037u.

(14) Zhou, J. J.; Feng, W. X.; Liu, C.-C.; Guan, S.; Yao, Y. G. *Nano Lett.* **2014**, *14*, 4767−4771.

(15) Ma, Y. D.; Dai, Y.; Yu, L.; Niu, C. W.; Huang, B. B. *New J. Phys.* **2013**, *15*, 073008.

(16) Wang, Z. F.; Chen, L.; Liu, F. *Nano Lett.* **2014**, *14*, 2879−2883.

(17) Xu, Y.; Yan, B. H.; Zhang, H.-J.; Wang, J.; Xu, G.; Tang, P.; Duan, W. H.; Zhang, S.-C. *Phys. Rev. Lett.* **2013**, *111*, 136804.

(18) Kou, L. Z.; Wu, S.-C.; Felser, C.; Frauenheim, T.; Chen, C. F.; Yan, B. H. *ACS Nano* **2014**, *8*, 10448–10454.

(19) Ma, Y. D.; Dai, Y.; Guo, M.; Niu, C. W.; Huang, B. B. *J. Phys. Chem. C* **2012**, *116*, 12977−12981.

(20) Si, C.; Liu, J. W.; Xu, Y.; Wu, J.; Gu, B.-L.; Duan, W. H. *Phys. Rev. B* **2014**, *89*, 115429.

(21) Qi, X-L.; Hughes, T. L.; Zhang, S.-C. *Phys. Rev. B* **2008**, *78*, 195424.

(22) Fu, L.; Kane, C. L. *Phys. Rev. B* **2007**, *76*, 045302.

(23) Bahramy, M. S.; Yang, B.-J.; Arita. R.; Nagaosa, N. *Nature Comm.* **2012**, *3*, 679.

(24) Wan, X.; Turner, A. M.; Vishwanath, A.; Savrasov, S. Y. *Phys. Rev. B* **2011**, *83*, 205101.

(25) Wang, J.; Chen, X.; Zhu, B-F.; Zhang, S.-C. *Phys. Rev. B* **2012**, *85*, 235131.

(26) Brüne, C.; Liu, C. X.; Novik, E. G.; Hankiewicz, E. M.; Buhmann, H.; Chen, Y. L.; Qi, X. L.; Shen, Z. X.; Zhang, S.-C.; Molenkamp. L. W. *Phys. Rev. Lett.* **2011**, *106*, 126803.

(27) Kuroda, K.; Ye, M.; Kimura, A.; Eremeev, S. V.; Krasovskii, E. E.; Chulkov, E. V.; Ueda, Y.; Miyamoto, K.; Okuda, T.; Shimada, K.; Namatame, H.; Taniguchi, M. *Phys. Rev. Lett.* **2010**, *105*, 146801.

(28) Bauer, E.; Hilscher, G.; Michor, H.; Paul, Ch.; Scheidt, E. W.; Gribanov, A.; Seropegin, Y.; Noël, H.; Sigrist, M.; Rogl, P. *Phys. Rev. Lett.* **2004**, *92*, 027003.

(29) Frigeri, P. A.; Agterberg, D. F.; Koga, A.; Sigrist, M. *Phys. Rev. Lett.* **2004**, *92*, 097001.

(30) Tanaka, Y.; Yokoyama, T.; Balatsky, A. V.; Nagaosa, N. *Phys. Rev. B* **2009**, *79*, 060505.

(31) König, M.; Wiedmann, S.; Brüne, C.; Roth, A.; Buhmann, H.; Molenkamp, L. W.; Qi, X. L.; Zhang, S.-C. *Science* **2007**, *318*, 766−770.

(32) Bernevig, B. A.; Zhang, S.-C. *Phys. Rev. Lett.* **2006**, *96*, 106802−106805.

(33) Kresse, G.; Furthmüller, J. *Comput. Mater. Sci.* **1996**, *6*, 15−50.

(34) Kresse, G.; Furthmüller, J. *Phys. Rev. B* **1996**, *54*, 11169−11186.

(35) Kresse, G.; Joubert, D. *Phys. Rev. B* **1999**, *59*, 1758−1775.

(36) Perdew, J. P.; Burke, K.; Ernzerhof, M. *Phys. Rev. Lett.* **1996**, *77*, 3865−3868.

(37) Monkhorst, H. J.; Pack, J. D. *Phys. Rev. B* **1976**, *13*, 5188−5192.

(38) Clark, S. J.; Segall, M. D.; Pickard, C. J.; Hasnip, P. J.; Probert, M. I. J.; Refson, K.; Payne, M. C. *Z. Kristallogr.* **2005**, *220*, 567−570.

(39) Refson, K.; Tulip, P. R.; Clark, S. J. *Phys. Rev. B* **2006**, *73*, 155114.

(40) Rashba, E. I. *Sov. Phys. Solid State* **1960**, *2*, 1109−1122.

(41) Zhang, H.; Liu, C.-X.; Qi, X.-L.; Dai, X.; Fang, Z.; Zhang, S.-C. *Nature Phys.* **2009**, *5*, 438−442.

(42) Heyd, J.; Scuseria, G.; Ernzerhof, M. *J. Chem. Phys.* **2003**, *118*, 8207-8215.

(43) Kane, C. L.; Mele, E. J. *Phys. Rev. Lett.* **2005**, *95*, 226801.

(44) Liu, C.-C.; Feng, W. X.; Yao, Y. G. *Phys. Rev. Lett.* **2011**, *107*, 076802.